\documentclass[epj]{webofc}
\usepackage[varg]{txfonts}   
\begin{document}
\title{Phenomenology of minimal Z' models: \\from the LHC to the GUT scale \footnote{Presented at \emph{QCD@Work2016}, 27-30 June 2016, Martina Franca, Italy} }

\author{\firstname{Elena} \lastname{Accomando}\inst{1}\fnsep\thanks{\email{e.accomando@soton.ac.uk}} \and
        \firstname{Claudio} \lastname{Corian\`o}\inst{2}\fnsep\thanks{\email{claudio.coriano@le.infn.it}} \and
        \firstname{Luigi} \lastname{Delle Rose}\inst{1,3}\fnsep\thanks{Speaker. \email{l.delle-rose@soton.ac.uk}} \and
        \firstname{Juri} \lastname{Fiaschi}\inst{1}\fnsep\thanks{\email{juri.fiaschi@soton.ac.uk}} \and
        \firstname{Carlo} \lastname{Marzo}\inst{2}\fnsep\thanks{\email{carlo.marzo@le.infn.it}} \and
        \firstname{Stefano} \lastname{Moretti}\inst{1}\fnsep\thanks{\email{s.moretti@soton.ac.uk}}
}

\institute{School of Physics and Astronomy, University of Southampton, Highfield, Southampton SO17 1BJ, UK
\and
	           Dipartimento di Matematica e Fisica Ennio De Giorgi, Universit\`a del Salento and INFN-Lecce, \\ Via Arnesano, 73100 Lecce, IT 
\and
           Dept. of Particle Physics, Rutherford Appleton Laboratory, Chilton, Didcot, OX11 0QX, UK
          }

\abstract{%
We consider a class of minimal abelian extensions of the Standard Model with an extra neutral gauge boson $Z'$ at the TeV scale. In these scenarios an extended scalar sector and heavy right-handed neutrinos are naturally envisaged. We present some of their striking signatures at the Large Hadron Collider, the most interesting  
arising from a $Z'$ decaying to heavy neutrino pairs as well as a heavy scalar decaying to two Standard Model Higgses. 
Using renormalisation group methods, we characterise the high energy behaviours of these extensions and exploit the constraints imposed by the embedding into a wider GUT scenario.
}
\maketitle
\section{Introduction}
\label{intro}
In this work we address a class of minimal models which are the most economical Abelian and renormalisable extensions of the Standard Model (SM) with only few additional free parameters. 
If the extra $U(1)'$ gauge symmetry is broken at energies accessible at the Large Hadron Collider (LHC), the corresponding $Z'$ gauge boson could provide a variety of new signatures. For instance, in addition to the di-lepton channel, the augmented flavour sector allows for the possibility of a $Z'$ decaying into long-lived heavy neutrinos with very clear multi-leptonic final states. 
The model under study is determined by the (SM) gauge group augmented by a single Abelian gauge factor $U(1)'$ which can always be described by a linear combination of the hypercharge and of the $B-L$ quantum number, with coefficients given by the new Abelian gauge coupling $\tilde g$ and $g_1'$. We choose the $U(1)_{B-L}$ as a reference gauge group and we explore an entire class of minimal Abelian models through the ratio of the extra Abelian gauge couplings. 
We investigate an Abelian extension of the SM with the only minimal matter content, namely a Right-Handed (RH) neutrino, one for each generation, with $B - L = -1$ and singlet under the SM group. This is necessary to ensure the cancellation of the gauge and gravitational anomalies. In the scalar sector we introduce a complex scalar field $\chi$, with $B - L = 2$ and singlet under the SM group, to achieve the spontaneous breaking of the extra Abelian symmetry. Its vacuum expectation value $x$, which we choose in the TeV range, provides the mass to the $Z'$ boson and to the RH neutrinos. The latter acquire a Majorana mass through the Yukawa interactions, dynamically realising the type-I seesaw mechanism. \\
The interactions of the charged fields with the neutral gauge bosons are described by the covariant derivative $\mathcal D_\mu = \partial_\mu + i g_1 \, Y \, B_\mu + i ( \tilde g \, Y + g'_1 \, Y_{B-L} ) B'_\mu + \ldots$, 
where $B_\mu$ and $B'_\mu$ are the gauge fields of the $U(1)_Y$ and $U(1)_{B-L}$ gauge groups, while $g_1$, $Y$ and $g'_1$, $Y_{B-L}$ are the corresponding couplings and charges. The parameter $\tilde g$ represents the mixing between the two Abelian groups.
The scalar potential is given by $V(H,\chi) = m_1^2 H^\dag H + m_2^2 \, \chi^\dag \chi + \lambda_1 (H^\dag H)^2 + \lambda_2 (\chi^\dag \chi)^2 + \lambda_3 (H^\dag H)(\chi^\dag \chi)$, 
where $H$ is the SM Higgs doublet. After spontaneous symmetry breaking the masses of the physical scalars are denoted by $m_{H_{1,2}}$, where $m_{H_1}$ is identified with the SM-like Higgs boson, while  the mixing angle between the two degrees of freedom is denoted by $\alpha$.
The Yukawa Lagrangian is $\mathcal L_Y =  \mathcal L_Y^{SM}  - Y_\nu^{ij} \, \overline{L^i} \, \tilde H \, {\nu_R^j}   - Y_N^{ij} \, \overline{(\nu_R^i)^c} \, {\nu_R^j} \, \chi + \, h.c. $
where $\mathcal L_Y^{SM}$ is the SM contribution. The light physical neutrinos are given by a combination of the left-handed SM neutrinos and a highly suppressed sterile RH component, while the heavier ones, with mass $m_{\nu_h} \simeq \sqrt{2}\, x Y_N$,  are mostly RH.

\section{Constraints from EWPTs and LHC searches}
\label{sec-1}
\begin{figure}[h]
\centering
\includegraphics[width=4.cm,clip]{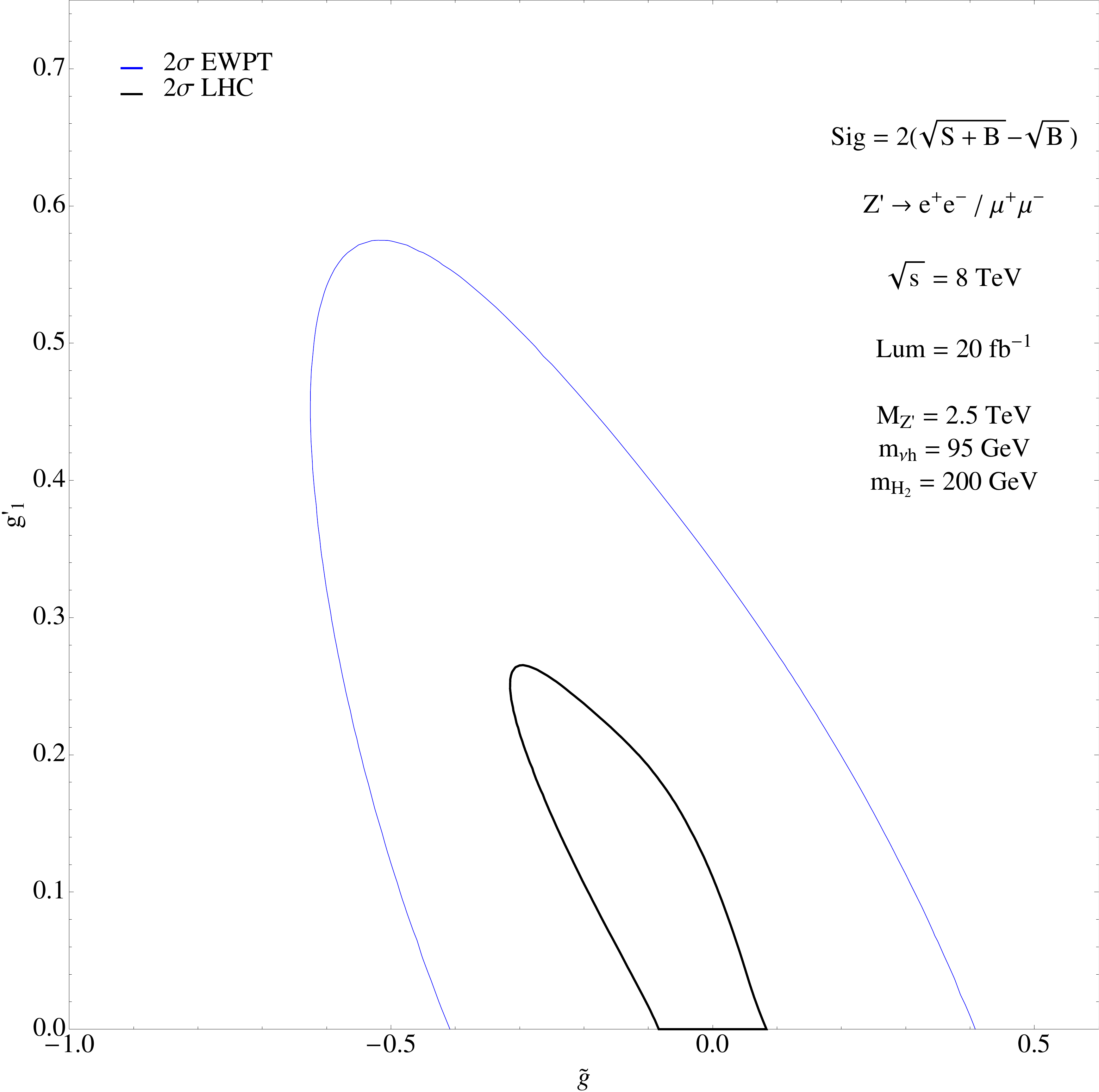} \qquad \qquad
\includegraphics[width=4.cm,clip]{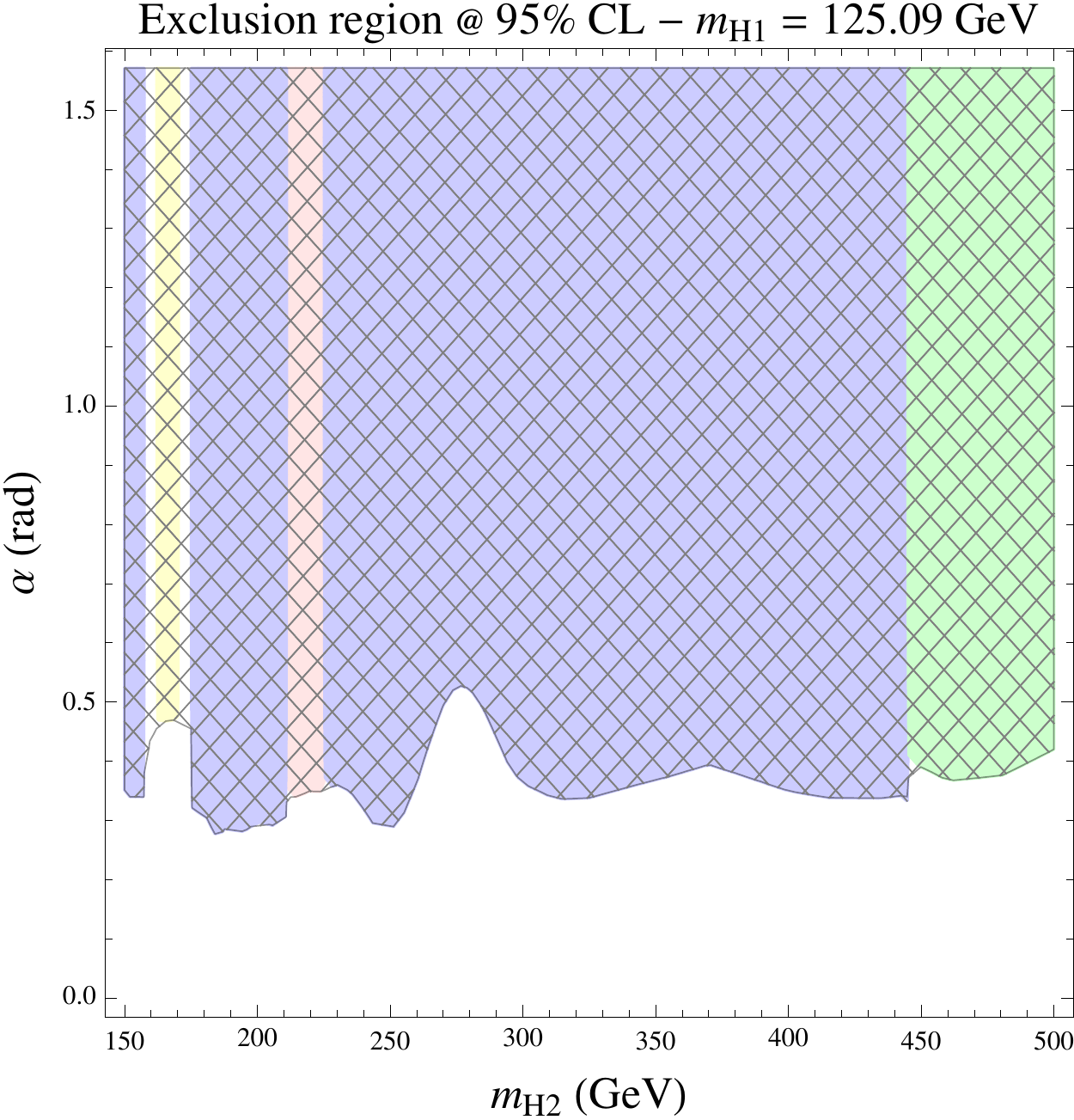}
\caption{(a) EWPTs and LHC bounds for $M_{Z'} = 2.5$ TeV. (b) Excluded region in the $(m_{H2}, \alpha)$ parameter space for $m_{\nu_h} = 95$ GeV and $M_{Z'} = 2$ TeV from \texttt{HiggsBounds}. The most sensitive exclusion channels are the four leptonic decay of two $Z$ bosons \cite{CMS:xwa} (red region), the full leptonic decay of two $W^\pm$ bosons \cite{CMS:bxa} (yellow region), the heavy Higgs decays into two $Z$s or $W^\pm$ s\cite{Khachatryan:2015cwa} (blue region) and  a combined search in five decay modes: $\gamma\gamma$, $ZZ$, $W^+W^-$, $\tau\tau$ and $bb$ \cite{CMS:aya} (green region). }
\label{fig-1}     
\end{figure}

Electroweak precision tests (EWPTs) from LEP2 data provide constraints on the $(\tilde g, g_1', M_{Z'})$ parameter space \cite{Salvioni:2009mt}. More stringent bounds can be extracted from the recent data of the Run-I of LHC at 8 TeV and $\mathcal L = 20$ fb$^{-1}$ using a signal-to-background analysis for the Drell-Yan (DY) channel (both electrons and muons are considered). We have included the Next-to-Next-to-Leading-Order QCD corrections to the DY computations through a k-factor and we have explicitly verified that the impact of the heavy neutrino mass and the parameters of the enlarged scalar potential have a marginal effect on the exclusion bounds (see also \cite{Klasen:2016qux}). We show in fig.~\ref{fig-1}(a) the $95\%$ Confidence Level (CL) exclusion limits from EWPTs and di-lepton analysis for $M_{Z'} = 2.5$ TeV. As mentioned above, the LHC analysis significantly improves the previous studies while comparable results are obtained only for high values of the $Z'$ mass. 
A similar study can be performed in the scalar sector. The present exclusion limits (combining LEP, Tevatron and LHC analyses) and the compatibility of the $Z'$ extension with the signal measurements of the discovered 125.09 GeV Higgs boson are taken into account, respectively, using \texttt{HiggsBounds} \cite{arXiv:0811.4169} and \texttt{HiggsSignals} \cite{Bechtle:2013xfa}. The \texttt{HiggsBounds} results are shown in the $(m_{H_2}, \alpha)$ parameter space in fig.~\ref{fig-1}(b). The most sensitive channels, covering almost all the considered mass interval, are $H \rightarrow W^{+}W^{-}$ and $H \rightarrow ZZ$ \cite{Khachatryan:2015cwa} (blue region). The requirement of the Higgs signal agreement at $2\sigma$ is found to provide a weaker bound with respect to the exclusion limits and therefore is not shown here.
\section{The high energy behaviour}
\begin{figure}[h]
\centering
\includegraphics[width=4.cm,clip]{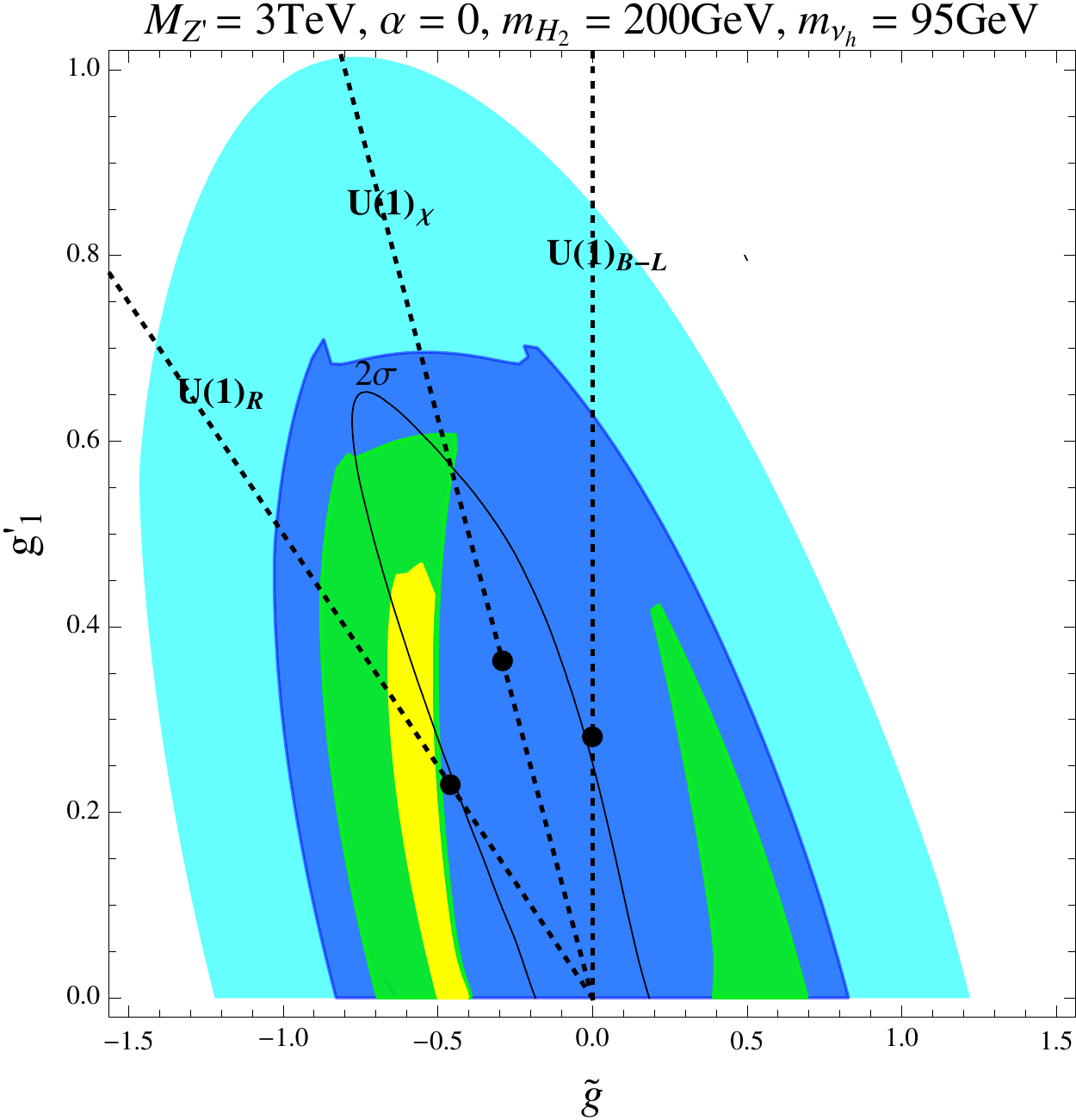}
\includegraphics[width=4.cm,clip]{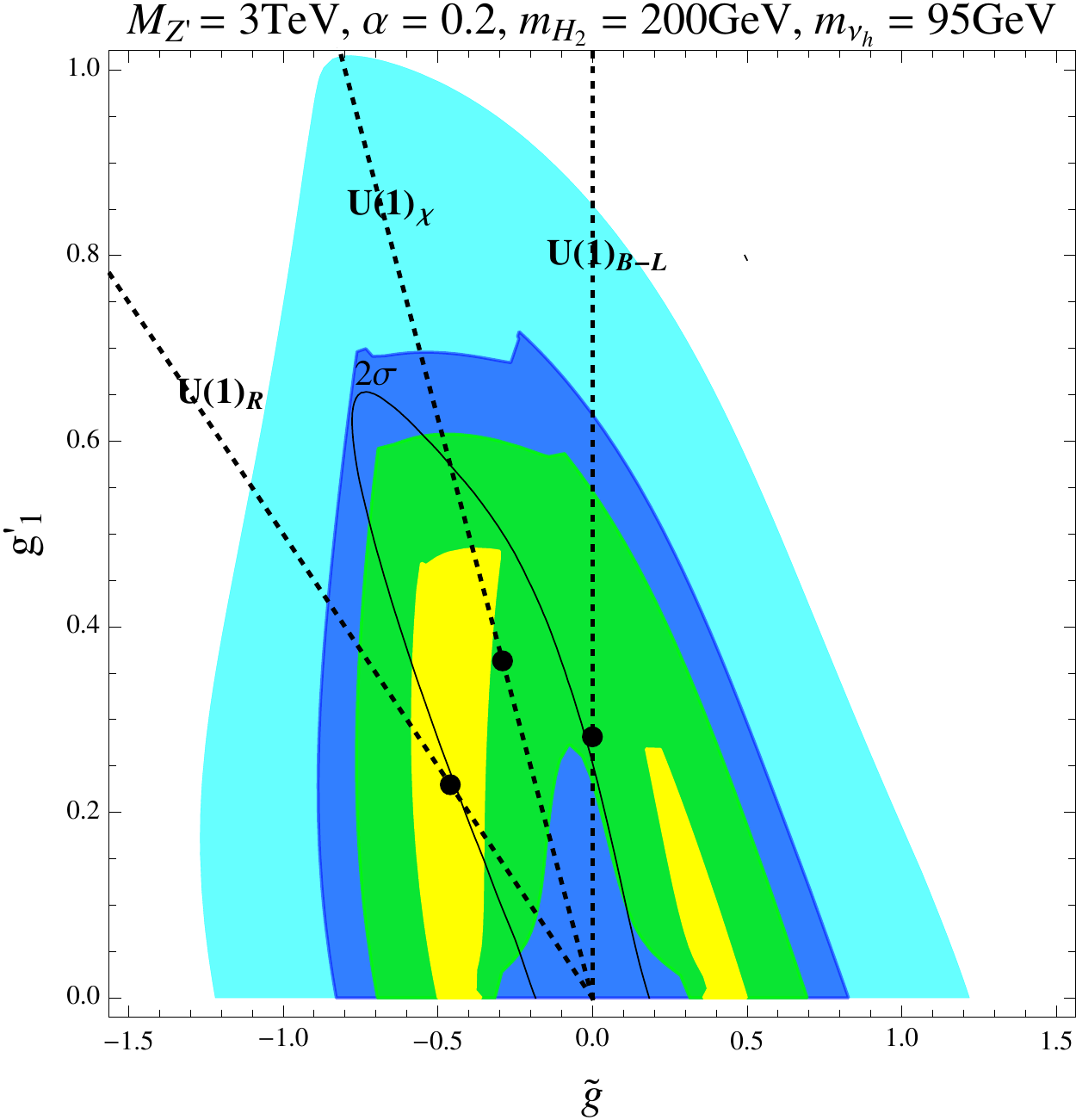}
\includegraphics[width=4.cm,clip]{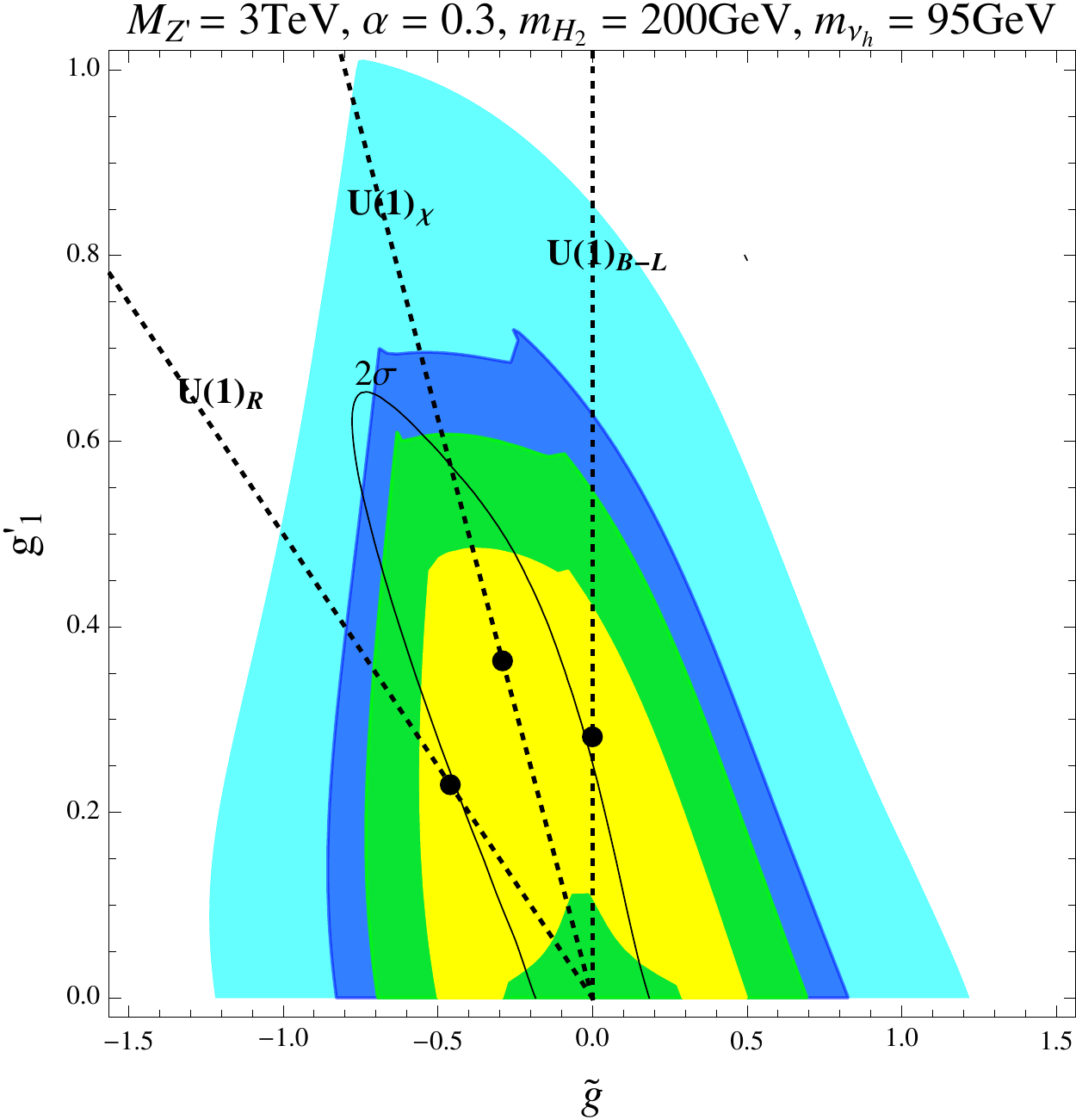}
\caption{Stability and perturbativity regions in the $(\tilde g, g_1')$ parameter space for three different values of the scalar mixing angle $\alpha$. The colour legend is depicted in fig.~\ref{fig-3}(b) }
\label{fig-2}     
\end{figure}

\begin{figure}[h]
\centering
\includegraphics[width=4.cm,clip]{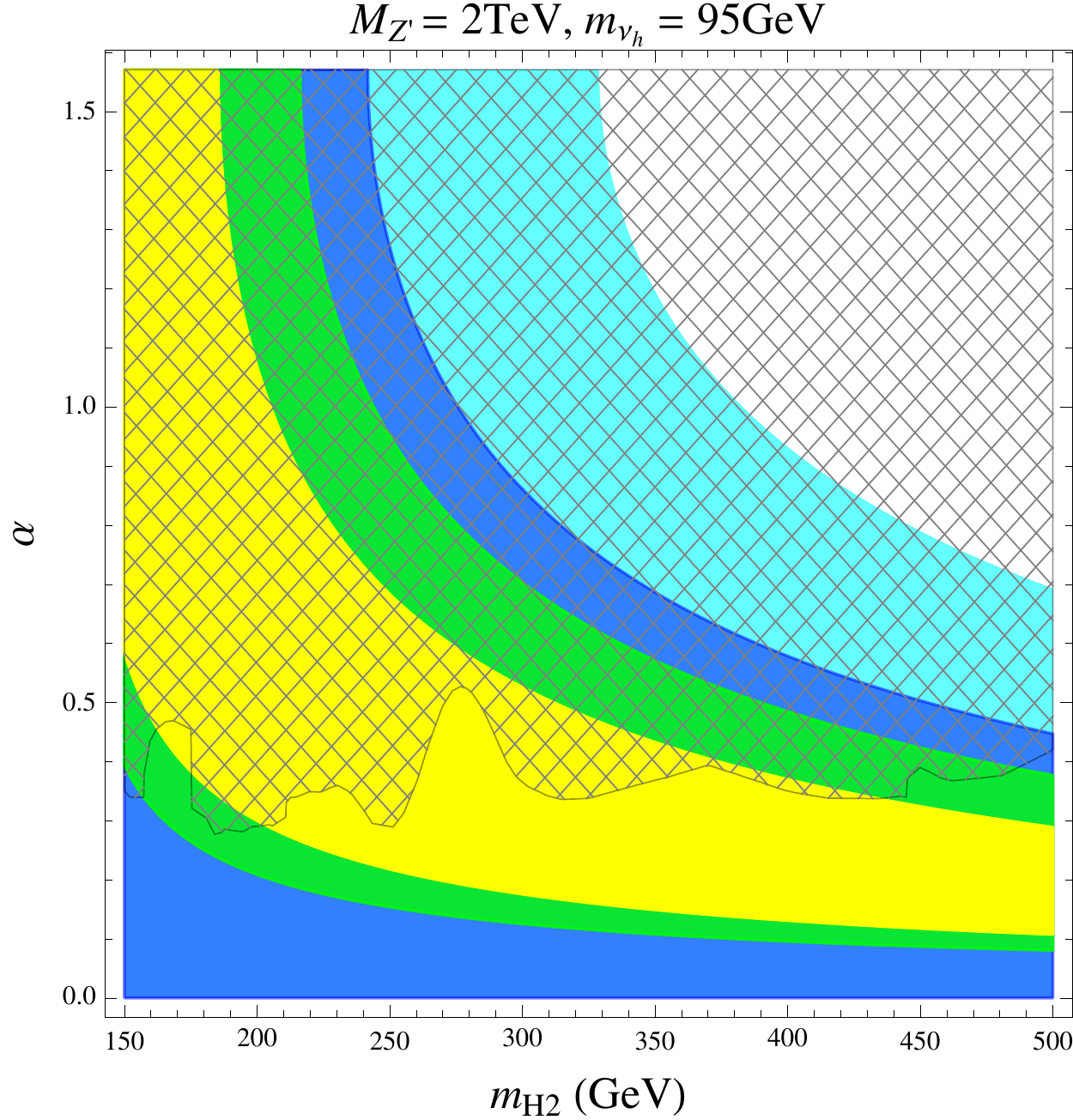} \qquad \qquad
\includegraphics[width=4.cm,clip]{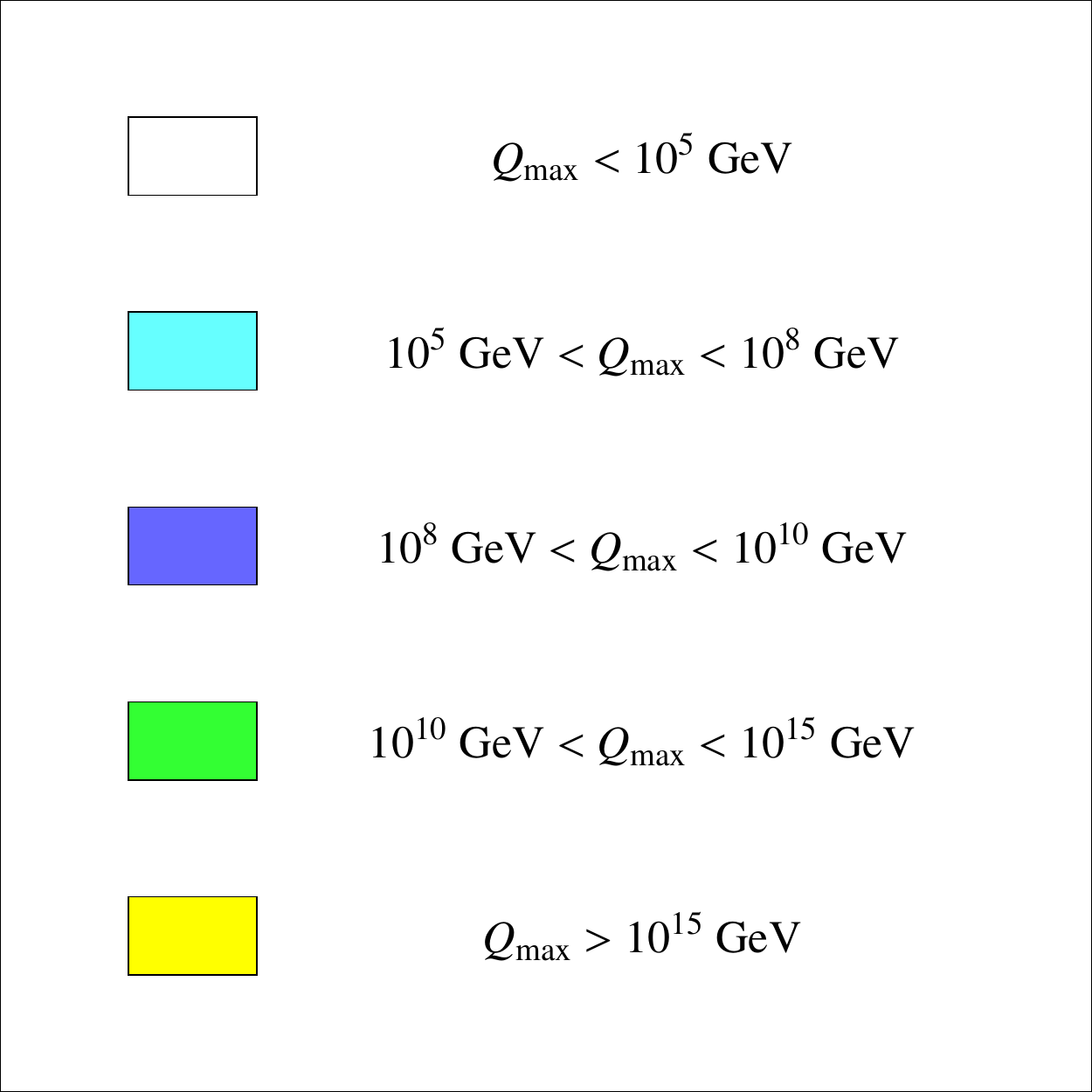}
\caption{(a) Stability and perturbativity regions in the $(m_{H_2}, \alpha)$ plane. The hatched area is excluded by \texttt{HiggsBounds}. (b) Legend of the different regions defined by the maximum scale $Q_{\rm max}$ up to which the model is stable and perturbative. }
\label{fig-3}     
\end{figure}

The tools of the Renormalisation Group (RG) theory can be exploited to establish a direct connection between accessible EW scale spectra and a potential underlying GUT structure. Therefore these methods can delineate the viable parameter space from both a phenomenological perspective and its theoretical consistency
and eventually guide the experimental investigations towards key analyses enabling one to make an assessment of the high energy structure of the model.
In this respect we explore the parameter space available at the EW scale under the requirement of perturbativity of the couplings and stability of the vacuum up to the GUT scale and beyond through the RG running.
We define $Q_{\rm max}$ as the maximum scale up to which the model possesses a stable scalar potential and perturbative couplings. The corresponding stability and perturbativity regions used in our analysis are depicted in fig.~\ref{fig-3}(b). For an easy comparison with the SM case we have introduced the region (denoted with the cyan colour) in which the stability and/or the perturbativity of these Abelian extensions is lost at a scale $Q_{\rm max}$ lower than the instability scale of the SM. The parameter space lying in this region compulsorily requires an ultraviolet completion, such a GUT scenario, with effects already visible at the $10^8$ GeV scale or even below.
On the other hand, the green and yellow regions (where stability and perturbativity are maintained up to the GUT scale and above) define the parameter space in which a more stable configuration is realised with respect to the SM, thus identifying them as viable extensions of the low energy theory. \\
We show in fig.~\ref{fig-2} the stability and perturbatibity regions in the parameter space defined by the two Abelian gauge couplings $\tilde g$,  $g'_1$ and for three different values of the mixing angle $\alpha$.
The results have been obtained for $M_{Z'} = 3$ TeV and $m_{\nu_h} = 95$ GeV. We refer to \cite{Coriano:2014mpa,Coriano:2015sea,Accomando:2016sge,DiChiara:2014wha} for the analysis of the impact of the $Z'$ and heavy neutrino masses.
The dashed lines represent three particular $U(1)'$ charge assignments, namely the pure $U(1)_{B-L}$, the $U(1)_\chi$ and the $U(1)_R$ extensions, while the sequential SM coincides with the $g'_1$ axis. The three dots represent the corresponding benchmark models usually addressed in the literature. 
From the three plots in fig.~\ref{fig-2} one can infer the impact of the scalar mixing angle $\alpha$ in improving the stability of the vacuum. 
This is, indeed, a common behaviour of scalar degrees of freedom. For $\alpha = 0$ the extra scalar $\chi$ is completely decoupled from the SM Higgs and the RG running of these $Z'$ extensions is similar to the SM if the additional $U(1)'$ gauge couplings are sufficiently small. For $\alpha > 0$ the two scalars interact with each other, thus preventing the instability of the vacuum up to GUT scale or above. \\
To highlight even more the role of the extended scalar sector we present in fig.~\ref{fig-2}(a) the same RG study in the $(m_{H_2}, \alpha)$ plane. Notice that the cyan region, in which high energy behaviour is worse than the SM case, is excluded for $m_{H_2} \lesssim 500$ GeV. Moreover, we remark that the impact of the one-loop matchings on the initial conditions and of the two-loop corrections on the $\beta$ functions is critical in the understanding of the high energy behaviour of the theory due to the sizeable corrections they provide on the scale at which the vacuum decay occurs \cite{Coriano:2015sea}.

\section{The LHC phenomenology}
\begin{figure}[h]
\centering
\includegraphics[width=4.2cm,clip]{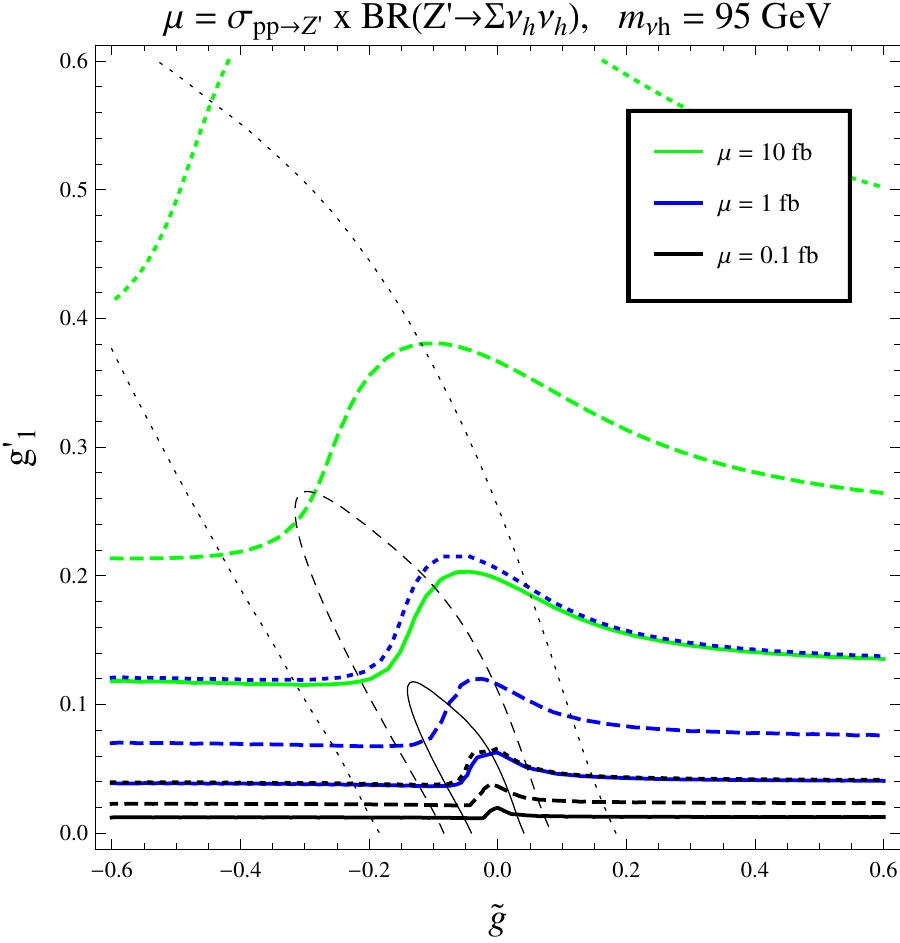} \qquad \qquad
\includegraphics[width=4.5cm,clip]{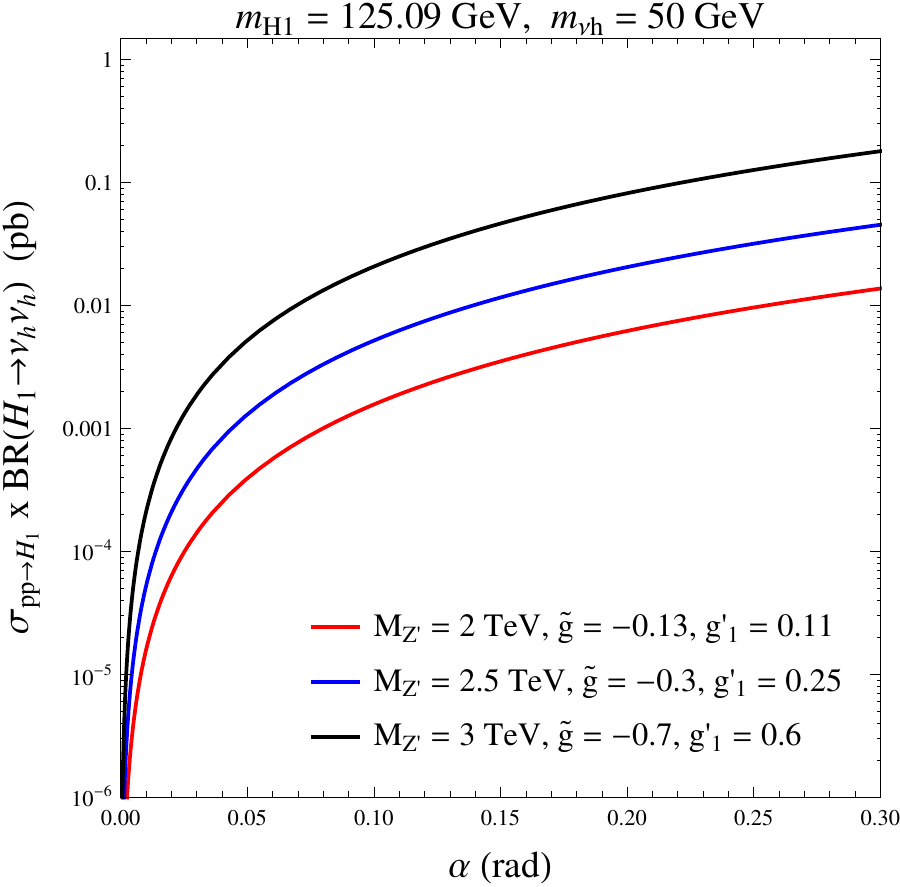}
\caption{(a) Cross section times BR for the process $pp \rightarrow Z' \rightarrow \nu_h \nu_h$ at the LHC with $\sqrt{s} = 13$ TeV in 
  the $(\tilde g, g_1')$ plane. Solid, dashed and dotted lines refer, respectively, to $M_{Z'} = $ 2, 2.5, 3 TeV.  (b) Cross section times BR for $pp \rightarrow H_1 \rightarrow \nu_h \nu_h$ at the LHC with $\sqrt{s} = 13$ TeV.}
\label{fig-4}     
\end{figure}

The decay mode hierarchy of the $Z'$ is strongly controlled by the gauge mixing $\tilde g$. For $\tilde g = 0$ the pure $B-L$ configuration is recovered. In this case the decay channel into charged leptons is the favourite one providing nearly the 40\% of the Branching Ratio (BR). For $\tilde g \gg g'_1$ we obtain the sequential $Z'$ in which the new gauge boson prefers the light quark decay mode with the corresponding BR reaching 60\%.
The BR into two heavy neutrinos shows an interesting variability with $g'_1$. Indeed, the partial width is found to be independent of $\tilde g$ and it is solely controlled by the Abelian coupling $g'_1$. Therefore, the corresponding BR ranges from zero, in the sequential limit, to $\sim 30\%$, its highest contribution, in the pure $B-L$ case.
The production of heavy neutrinos from $Z'$ decay, which we recall are required to cancel the anomalies associated to the new gauge group, 
is a hallmark of $U(1)'$ scenarios. The successive decays of the heavy neutrino may result in striking multi-lepton signature (see, for instance, \cite{Emam:2007dy} for the 2-lepton, \cite{Basso:2008iv} for the 3-lepton and  \cite{Huitu:2008gf,Khalil:2015naa} for the 4-lepton channel).
In this respect we show in fig.~\ref{fig-4}(a) the contour plots of the $\sigma \times \textrm{BR}$ for the production of heavy neutrinos from a decaying $Z'$ in the plane defined by the new Abelian gauge couplings. The cross section is computed for a Centre-of-Mass (CM) energy of 13 TeV for different values of $M_{Z'}$.
Notice that the $Z'$ on-shell production is characterised by a cross section up to $\sigma=5$ fb for $M_{Z'}=2$ TeV and up to $\sigma=10$ fb for $M_{Z'}=3$ TeV in the pure $B-L$ case, where $\tilde g = 0$. Changing $\tilde{g}$ modifies the $Z'$ coupling to the light quarks and may also allow more sizeable cross sections. For instance, we approach $\sigma=100$ fb for $M_{Z'}=3$ TeV and $\tilde{g}=-0.6$.

The possibility to explore different $Z'$ model scenarios, provided by a non vanishing gauge mixing, opens new decay channels of the $Z'$ into SM bosons, which are absent in the pure $B-L$ case, namely, $Z'\to W^+W^-$, $Z\,H_1$ and $Z\,H_2$.
Among the three, the decays into $W^+W^-$ and  $Z\,H_1$ represent the main channels, with BR $\sim 2\%$, 
regardless of the value of the scalar mixing angle $\alpha$ (at least in its allowed interval) with  kinematics accounting for the main differences. 
On the other hand, the BR of the $Z \, H_2$ channel does not exceed the 0.1\% value.

The existence of an augmented scalar sector of a $U(1)'$ origin offers the chance to explore completely novel signatures with respect to many beyond SM extensions. Among the others, the heavy Higgs decay into heavy neutrinos, $H_2 \rightarrow \nu_h \nu_h$, represents a hallmark of $U(1)'$ scenarios. Moreover, a non-zero scalar mixing angle $\alpha$ provides, if kinematically allowed $m_{H_2} > 2 \,m_{H_1}$, the decay of the heavier scalar into two light ones, $H_2 \rightarrow H_1 H_1$, which is a unique opportunity to investigate the mechanism of spontaneous symmetry breaking.
As already stated above, the $\alpha$ parameter plays a key role in the phenomenology of scalars, scaling all $H_1$ ($H_2$) couplings with SM-like particles by $\cos(\alpha)$ ($\sin(\alpha)$) and by $\sin(\alpha)$ ($\cos(\alpha)$) when involving particles peculiar of the $U(1)'$ extension, namely the $Z'$ and heavy neutrinos.
When $m_{H_1} > 2 m_{\nu_h}$, a new interesting channel become accessible to the SM-like Higgs boson, $H_1\to \nu_h \nu_h$, otherwise it behaves exactly as the SM Higgs, with the same BRs and a total width rescaled by a factor of $\cos^2 \alpha$. The BR into two heavy neutrinos ranges from 0.1\% to 1\% becoming comparable to, or even exceeding, the $\gamma\gamma$ mode of the SM Higgs.
We show in fig.~\ref{fig-4}(b) the $\sigma \times \textrm{BR}$ as a function of the mixing angle for the process $pp \rightarrow H_1 \rightarrow \nu_h \nu_h$ at the LHC with $13$ TeV, for $m_{\nu_h} = 50$ GeV and for three different benchmark points defined by $M_{Z'}, \tilde g$ and $g_1'$. We have only considered the gluon fusion channel which can provide $\sigma \times \textrm{BR} \sim 100$ fb. The $H_1$ production cross section scales with a factor of $\cos^2 \alpha$ with respect to the SM case, reproduced by $\alpha = 0$. In such case, $\sigma({gg \rightarrow H_1}) = 44.08$ pb \cite{LHCHXSWG}.
The behaviour of the three benchmark points is related to the structure of the $H_1 \nu_h \nu_h$ vertex which is proportional to $\sin \alpha \, ( m_{\nu_h} / x) \sim \sin \alpha \, g'_1 ( m_{\nu_h} / M_{Z'})$. For a given $m_{\nu_h}$, the interaction of the SM-like Higgs with the heavy neutrinos gets stronger by growing the ratio $g'_1/M_{Z'}$. Taking into account the LHC limits on the Abelian gauge couplings, which are less constraining for higher $Z'$ masses, we find a bigger ratio for $M_{Z'} = 3$ TeV, in which case $g'_1$ is allowed to vary up to 0.6 for $\tilde g = -0.7$.

\section*{Acknowledgments}
E.A., L.D.R., J.F. and S.M. are supported in part through the NExT Institute. C.C. thanks The
Leverhulme Trust for a Visiting Professorship to the University of Southampton, where
part of his work was carried out. The work of L.D.R. has also been supported by the "Angelo
Della Riccia" foundation and a COFUND/STFC Rutherford International Fellowship.


\begin{thebibliography}{}

\bibitem{Salvioni:2009mt}
E.~Salvioni, G.~Villadoro, and F.~Zwirner, {\it {Minimal Z-prime models:
  Present bounds and early LHC reach}},  {\em JHEP} {\bf 11} (2009) 068,
  arXiv:0909.1320.

\bibitem{Klasen:2016qux}
M.~Klasen, F.~Lyonnet and F.~S.~Queiroz, {\it {NLO+NLL Collider Bounds, Dirac Fermion and Scalar Dark Matter in the B-L Model}},
  arXiv:1607.06468.

\bibitem{arXiv:0811.4169}
P.~Bechtle, O.~Brein, S.~Heinemeyer, G.~Weiglein, and K.~E. Williams, {\it
  {HiggsBounds: Confronting Arbitrary Higgs Sectors with Exclusion Bounds from
  LEP and the Tevatron}},  {\em Comput. Phys. Commun.} {\bf 181} (2010)
  138--167, arXiv:0811.4169.

\bibitem{Bechtle:2013xfa}
P.~Bechtle, S.~Heinemeyer, O.~Stal, T.~Stefaniak, and G.~Weiglein, {\it
  {$HiggsSignals$: Confronting arbitrary Higgs sectors with measurements at the
  Tevatron and the LHC}},  {\em Eur. Phys. J.} {\bf C74} (2014), no.~2 2711, arXiv:1305.1933.

\bibitem{Khachatryan:2015cwa}
{\bf CMS} Collaboration, V.~Khachatryan {\em et.~al.}, {\it {Search for a Higgs
  Boson in the Mass Range from 145 to 1000 GeV Decaying to a Pair of W or Z
  Bosons}},  {\em JHEP} {\bf 10} (2015) 144,
  arXiv:1504.00936.

\bibitem{CMS:xwa}
{\bf CMS} Collaboration, {\it {Properties of the Higgs-like boson in the decay
  H to ZZ to 4l in pp collisions at sqrt s =7 and 8 TeV}}, CMS-PAS-HIG-13-002.

\bibitem{CMS:bxa}
{\bf CMS} Collaboration, {\it {Update on the search for the standard model
  Higgs boson in pp collisions at the LHC decaying to W + W in the fully
  leptonic final state}}, CMS-PAS-HIG-13-003.

\bibitem{CMS:aya}
{\bf CMS} Collaboration, {\it {Combination of standard model Higgs boson
  searches and measurements of the properties of the new boson with a mass near
  125 GeV}}, CMS-PAS-HIG-12-045.


\bibitem{Coriano:2014mpa}
C.~Coriano, L.~Delle~Rose, and C.~Marzo, {\it {Vacuum Stability in U(1)-Prime
  Extensions of the Standard Model with TeV Scale Right Handed Neutrinos}},
  {\em Phys. Lett.} {\bf B738} (2014) 13--19, arXiv:1407.8539.
  
\bibitem{Coriano:2015sea}
C.~Coriano, L.~Delle~Rose, and C.~Marzo, {\it {Constraints on abelian
  extensions of the Standard Model from two-loop vacuum stability and
  $U(1)_{B-L}$}},  {\em JHEP} {\bf 02} (2016) 135, arXiv:1510.02379.

\bibitem{Accomando:2016sge}
  E.~Accomando, C.~Coriano, L.~Delle Rose, J.~Fiaschi, C.~Marzo and S.~Moretti,
  {\it {Z$^{'}$, Higgses and heavy neutrinos in U(1)$^{'}$ models: from the LHC to the GUT scale}},
  {\em JHEP} {\bf 1607} (2016) 086, arXiv:1605.02910.

\bibitem{DiChiara:2014wha}
S.~Di Chiara, V.~Keus and O.~Lebedev, {\it {Stabilizing the Higgs potential with a Z$'$}},
 {\em Phys. Lett.}  {\bf B744} (2015) 59, arXiv:1412.7036.
  
\bibitem{Emam:2007dy}
W.~Emam and S.~Khalil, {\it {Higgs and Z-prime phenomenology in B-L extension
  of the standard model at LHC}},  {\em Eur. Phys. J.} {\bf C52} (2007)
  625--633, arXiv:0704.1395.

\bibitem{Basso:2008iv}
L.~Basso, A.~Belyaev, S.~Moretti, and C.~H. Shepherd-Themistocleous, {\it
  {Phenomenology of the minimal B-L extension of the Standard model: Z' and
  neutrinos}},  {\em Phys. Rev.} {\bf D80} (2009) 055030,
  arXiv:0812.4313.

\bibitem{Huitu:2008gf}
K.~Huitu, S.~Khalil, H.~Okada, and S.~K. Rai, {\it {Signatures for right-handed
  neutrinos at the Large Hadron Collider}},  {\em Phys. Rev. Lett.} {\bf 101}
  (2008) 181802, arXiv:0803.2799.

\bibitem{Khalil:2015naa}
S.~Khalil and S.~Moretti, {\it {The $B-L$ Supersymmetric Standard Model with
  Inverse Seesaw at the Large Hadron Collider}},
 arXiv:1503.08162.

\bibitem{LHCHXSWG}
{LHC Higgs Cross Section Working Group}, \\{\it
  {https://twiki.cern.ch/twiki/bin/view/LHCPhysics/CERNYellowReportPageAt13TeV}}.

\end{thebibliography}
\end{document}